# A Method of Evaluating Effect of QoS Degradation on Multidimensional QoE of Web Service with ISO - based Usability

Daisuke Yamauchi[1] and Yoshihiro Ito[2]

[1] Nagoya Institute of Technology, Japan
yamauchi@en.nitech.ac.jp
[2] Nagoya Institute of Technology, Japan
yoshi@nitech.ac.jp

*ABSTRACT*

*This paper studies a method of investigating effect of IP performance (QoS) degradation on quality of experience (QoE) for a Web service; it considers the usability based on the ISO 9241-11 as multidimensional QoE of a Web service (QoE-Web) and the QoS parameters standardized by the IETF. Moreover, the paper tackles clarification of the relationship between ISO-based QoE-Web and IETF-based QoS by the multiple regression analysis. The experiment is intended for the two actual Japanese online shopping services and utilizes 35 subjects. From the results, the paper quantitatively discusses how the QoE-Web deteriorates owing to the QoS degradation and shows that it is appropriate to evaluate the proposed method.*

## 1. INTRODUCTION

Some of Web services have recently become indispensable for our life while we can utilize many Web services over the Internet, for instance, Web mapping services [1], online shopping services [2], and so forth. As a Web service becomes more essential for us, its higher quality is required.

Since a Web service is provided over the Internet, quality of the service can deteriorate because of performance degradation of the Internet. Consequently, it is significant to study effect of network performance degradation on quality of a Web service. To accomplish this, we first of all clarify both quality of a Web service and performance of the Internet.

Firstly, quality of a Web service can be assessed from many points of view. Among of them, it is often required to assess the quality from a user's point of view (user-centric quality), which is also referred to as *Quality of Experience (QoE)*. Note that, indeed QoE is a kind of *quality of service (QoS)*, this paper refers to (lower-level) QoS except for QoE as QoS for the sake of convenience. International Telecommunication Union Telecommunication Standardization Sector (ITU-T) considers QoE for a Web service in G.1010 [3] and G.1030 [4]. ITU-T SG 12 is now studying recommendation for QoE of Web-browsing and the recommendation will be published as G.1031. However, they chiefly treat only a single measure, that is, *Mean Opinion Score (MOS)*, as QoE and are not necessarily appropriate to current diverse complicated Web services.

Alternatively, we can assess QoE for a Web service (Web-QoE) with usability, which is defined by Nielsen [5] or the International Organization for Standardization (ISO) [6]. Usability of a Web service is called as Web usability [7]. ISO has standardized that the usability has three aspects while [5] mentions that the usability is assessed from the five points of view. Since the usability can quantify QoE multidimensionally, it is more suitable than MOS as Web-QoE. This paper therefore considers Web usability defined by ISO as Web-QoE.

Secondly, as performance evaluation of the Internet, we can utilize the framework for the Internet Protocol (IP) performance metric [8], which has been standardized by the Internet Protocol Performance Metric Working Group (IPPM WG) of the Internet Engineering Task Force (IETF). In this framework, a lot of measures of IP performance evaluation in RFCs are standardized.

Many researches treat Web usability. For example, [9] studies how older adults interact with a Web service. Reference [10] shows both international differences and effects of high-end graphical enhancements on perceived usability of a Web service. In [11], crucial Web usability factors of Web services for students are studied from 36 industries. However, a lot of studies about the Web usability regard their network as an ideal one.

This paper studies a method of clarifying effect of IP performance degradation on multidimensional QoE of a Web service with the usability. In our experiment, we adopt an online shopping service as a target Web service. We would like to tackle clarifying relationship between QoE-Web based on ISO's usability and IP performance metrics defined by IETF. The remainder of this article is organized as follows. Section 2 shows the related works. Section 3, we introduce a Web usability defined by ISO. Section 4 describes QoS evaluation. Sections 5 and 6 depict our experiments and their results, respectively. Finally, we conclude our research in Section 7.

## 2. RELATED WORKS

We find some papers concerning qualitative relationship between QoE and QoS for Web services [12] [13] [14]. For example, [12] studies generic exponential relationship between QoE (MOS) and QoS for Web browsing. In [13], influence of waiting time on MOS for Web-based services is discussed. Reference [14] investigates how bandwidth and resulting waiting time affect MOS for Web browsing. On the other hand, [15] studies the effect of routing and TCP variants on the HTTP and FTP performance. However, almost all the researches in the field of networks treat the standards defined by the ITU or the IETF and scarcely consider the ISO-based usability as a measure of QoE in Web services. Therefore, little is known about the effect of the IP performance degradation on the ISO's Web usability.

## 3. WEB USABILITY

### 3.1. Usability

ISO has provided some international standards concerning usability as follows. ISO 9241-11 [6] defines usability. Based on this standard, ISO 13407 [16] treats technical human factors and ergonomics issues in the design process. ISO 9126 [17] classifies software quality in a structured set of characteristics and sub-characteristics; it uses usability as one of attributes.

### 3.2. ISO 9241-11

ISO 9241-11 defines usability of a visual display terminal (VDT). In this standard, usability indicates degree to which a product can be used by specified users to achieve specified goals in a specified context of use; it consists of three attributes: effectiveness, efficiency and satisfaction. In [6], they are defines as follows. The effectiveness means accuracy and completeness with which users achieve specified goals. The efficiency indicates resources expended in relation to the accuracy and completeness with which users achieve goals. The satisfaction depicts freedom from discomfort, and positive attitudes towards the use of the product.

In ISO 9241-11, when measuring usability, the following information is required: a description of the intended goals, a description of the components of the context of use, which includes users, tasks, equipment and environments, and target values of effectiveness, efficiency, and satisfaction. Moreover, we must define at least one measure for each of effectiveness, efficiency and satisfaction. However, because the relative importance of attributes of usability depends on the context of use and the purposes, ISO 9241-11 shows no concrete rule for how measures should be chosen.

## 4. QoS evaluation

The IPPM WG has been developing a series of standard measures that can be applied to the quality, performance, and reliability of the Internet; it has issued documents on the standards as RFC. For example, RFC 2330 [8] defines a general framework for particular metrics to be developed by IPPM WG. RFC 2678 [12] defines metrics for connectivity between a pair of Internet hosts. RFC 2680 [18] handles a metric for one-way packet loss across paths. RFC 2679 [19] and RFC 2681 [20] treat a metric for one-way delay of packets and that for round trip delay, respectively. RFC 3393 [21] refers to a metric for variation in delay of packets across paths. RFC 6349 [22] describes a methodology for measuring end-to-end TCP Throughput. RFC 4656 [23] and RFC 5357 [24] treat an active measurement protocol.

## 5. Experiments

In our experiments, we assess Web usability for actual Web services in accordance with ISO 9241-11 as we evaluate IP performance. As the first step of our research, we adopt online shopping services as target Web services. This section describes a usability requirements specification of our experiments in accordance with ISO 9241-11.

### 5.1. Name and goal of the Web service

We consider usability for the following two online shopping Web services in Japan. They are the first and second ranking online shopping services in Japan [25]. One is a huge single worldwide store (We refer to ServiceA in the rest) and the other is a shopping mall and an aggregate of over 37 thousands of stores (We refer to ServiceB.)

The goal of our subjects (users) is to buy some designated commodities that they want. However the subjects cannot always find their ideal commodities. Thus we set some conditions for each commodity. Even if the subjects cannot finish their task (goal), they can accomplish some of the conditions (sub-goals).

### 5.2. Context of use in our experiments

#### 5.2.1. Users (subjects)

Users are male and female in twenties. The number of them is 35.

#### 5.2.2. Task

For the above-mentioned two Web services, users are tasked with buying the following five usual commodities designated by the experimenter: a personal computer, a bicycle, a (computer) mouse, a garbage can and a Digital Versatile Disc (DVD) movie. In addition to this, the experimenter imposes six conditions for every commodity on the subjects. When the users want to buy the commodities, they cannot always fulfil all the conditions. We therefore give three levels of priority on them: ``High priority'', ``Middle priority'' and ``Low priority''. The users are explained that the ratio of the priority is 6:3:1; they put more effort into a task that has higher priority.

#### 5.2.3. Environment

Figure 1 depicts our experimental configuration. In this configuration, a network emulator connects a Web client is connected with the Internet via a network emulator that can delay packets and randomly drop them at a constant rate.

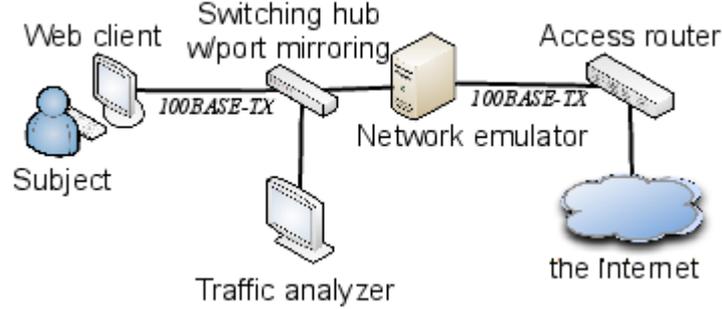

Figure 1. Experimental configuration.

By controlling round trip delay and packet loss rate, we change our experimental environment. We consider three combinations of a value of the round trip delay and that of the packet loss rate as shown in Table 1. For the convenience sake, we give numbers to the environments from 1 to 5.

Table 1. Five experimental environments.

| Experimental environment | Added round trip delay (ms) | Added packet loss rate (%) |
|---|---|---|
| 1 | 0 | 0 |
| 2 | 150 | 0 |
| 3 | 0 | 5 |
| 4 | 150 | 5 |
| 5 | 200 | 10 |

The network emulator also measures traffic between the Web client and the Internet for IP performance evaluation.

### 5.2.4. Equipment

We utilize Firefox 3.6 [26] and Dummynet [27] as the Web client and the network emulator, respectively; we adopt Tcpdump [28] to evaluate the IP performance.

### 5.3. Measures of Web usability

#### 5.3.1. Effectiveness

We consider the proportion of achieved conditions of our tasks to all the conditions as a measure of the effectiveness. We weight the proportion according to its priority and calculate an average of all users. The effectiveness $E$ is defined as

$$E = \frac{1}{N}\sum_{i=1}^{N}\sum_{p=H,M,L} w_p \frac{c_p^i}{C_p} \qquad (1)$$

where $N$ is the number of users, and $C_H$, $c_H^i$ and $w_H$ mean the number of all the conditions with high priority, that of achieved conditions among them by the $i$-th user and the weight of High priority, respectively. Similarly, ($C_M$, $c_M^i$, $w_M$) and ($C_L$, $c_L^i$, $w_L$) are defined for

Middle priority and for Low priority, respectively. According to the priority defined in 4.2.2, we set $w_H$ =0.6, $w_M$ =0.3 and $w_L$ =0.1.

### 5.3.2. Efficiency

A measure of the efficiency is considered to be the effectiveness for workload to finish a task per user. The workload for accomplishing a task closely relates to the time to consume in moving mouse, typing keyboard, and so on [29]. Thus, we first define the workload *W* as

$$W = \frac{1}{N}(I_s s_i + I_m m_i + I_b b_i + I_k k_i) \quad (2)$$

where, for the *i*-th user, $s_i$, $m_i$, $b_i$ and $k_i$ indicate the number of the spins of a mouse wheel, the distance of mouse movement, the number of mouse clicks and the number of keystrokes, respectively. On the other hand, $I_s$, $I_m$, $I_b$ and $I_k$ mean the average wheel spins per unit time, the average velocity of mouse, the average number of clicks per unit time and the typing speed, respectively. By using the coefficients $I_s$, $I_m$, $I_b$ and $I_k$, we can calculate the time consumed by the user to accomplish a task. As a result of our prior experiments, we get $I_s$ =100, $I_m$ =10000, $I_b$ =20 and $I_k$ =20. Then, we define the efficiency *H* as

$$H = \frac{E}{W} \quad (3)$$

### 5.3.3. Satisfaction

We measure the satisfaction by adopting psychological methods as follows. We first use the rating scale method [30] with seven levels. In this method, the users are to rate each stimulus (Web service) with respect to their satisfaction. We instruct the users to rate on a scale from 1 to 7. If a Web service is rated 7 by a user, we can consider that the user perfectly satisfies the service. Next, the satisfaction is calculated from the result of the rating scale method with the law of categorical judgment [30]. The law can translate an ordinal scale [30] measured by the rating scale method into the psychological interval scales [30]. It should be noted that a scale obtained by the law is an interval scale and has no origin. We therefore define the smallest value among obtained ones for stimuli as the origin.

## 5.4. IP performance metrics

For the first step of our research, this paper considers the following metrics of IP performance evaluation. First, we consider the round trip delay defined in RFC 2681 and the packet loss metric defined in RFC 2680. Second, since we try to use statistics concerning TCP, we treat some metrics defined in RFC 6349: average TCP segment size, number of packets transmitted or received per unit time, amount of transmitted or received data per unit time, number of retransmitted packets and number of retransmitted byte.

## 6. RESULTS AND CONSIDERATIONS

## 6.1. Results of QoS evaluation

At first we show the results of QoS evaluation in Fig. 2 through Fig. 9. In these figures, the abscissa indicates our experimental environment described in the previous section; we also plot 95% confidence intervals.

Figures 2 and 3 show the measured round trip delays. The former plots the results which were measured with the TCP segments for connection establishment, and the latter indicates those for all TCP segments.

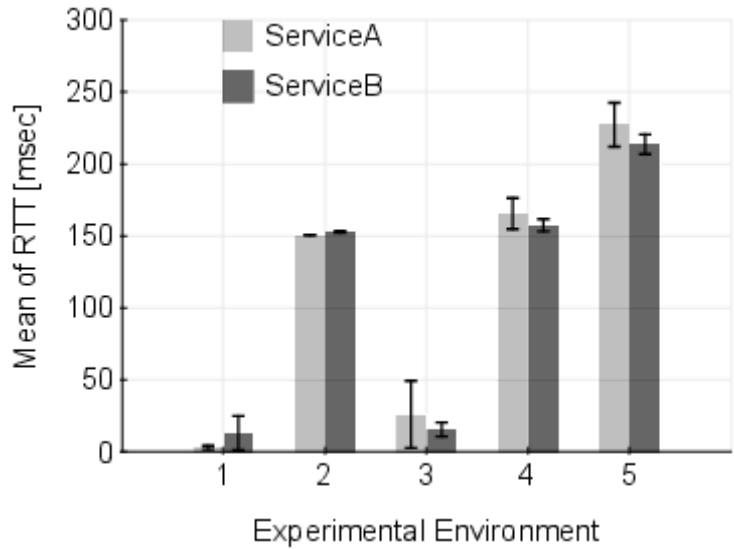

Figure 2. Mean of round trip delay.

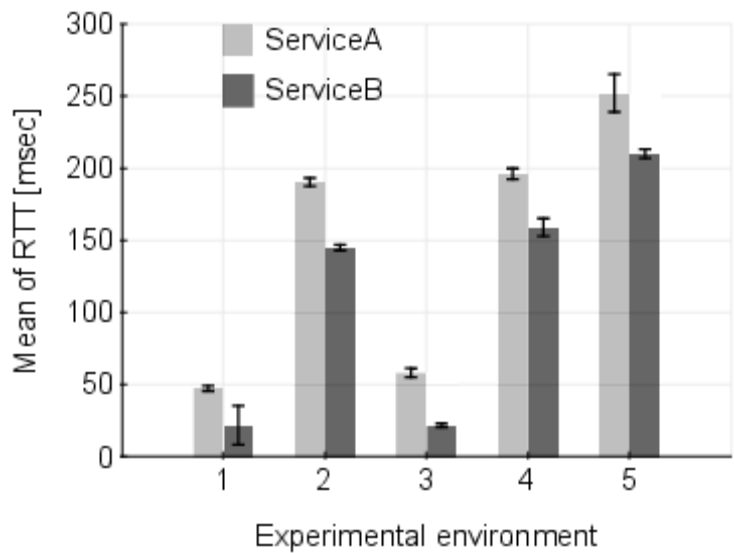

Figure 3. Mean of round trip delay for all TCP segments.

From Fig. 2, we see that the measured transmission delays are about the same as the values added by the network emulator; there is no significant difference between RTTs for the two services. On the other hand, Fig. 3 shows that the mean of the actual RTTs for ServiceA is larger than those for ServiceB by 30 milliseconds.

Figure 4 displays the mean of TCP segment length, and shows that mean of TCP segment for ServiceA is longer than that for ServiceB by 200 bytes.

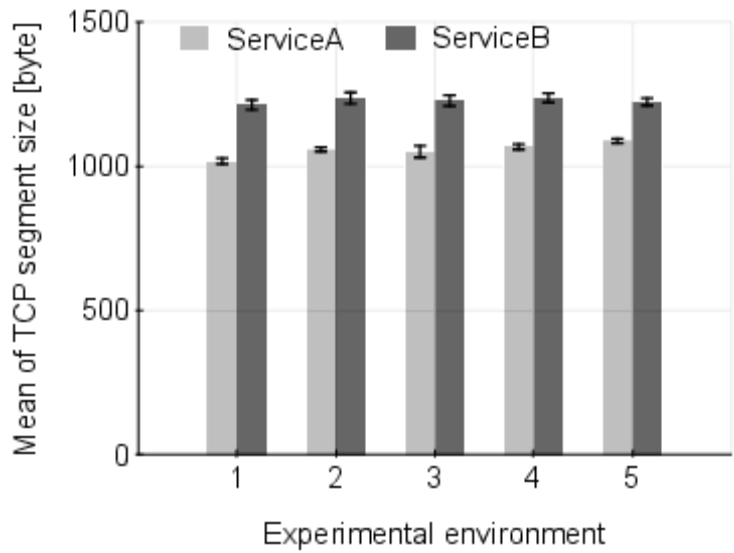

Figure 4. Mean of TCP segment length.

We display the number of transmitted packets per unit time and the amount of transmitted data per unit time in Fig. 5 and Fig. 6, respectively. From these figures, we find that the amount of transmitted data of ServiceB is more than that of ServiceA. Moreover, while the amount of ServiceB decreased because of QoS degradation, that of ServiceA did not.

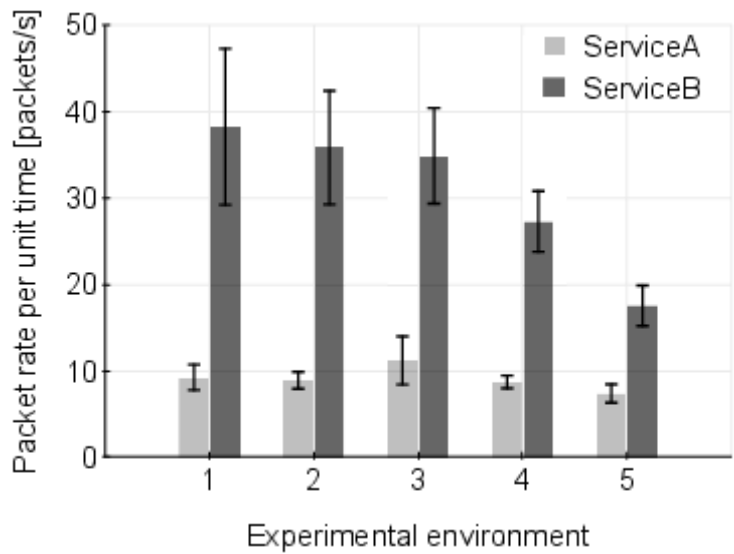

Figure 5. Number of the transmitted packets per unit time.

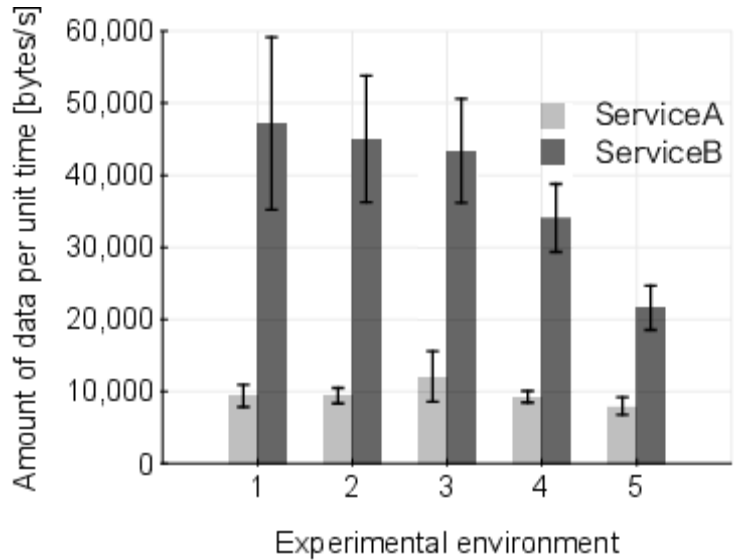

Figure 6. Amount of the transmitted data per unit time.

Figures 7 and 8 plot the number of retransmitted TCP segments and the amount of retransmitted data, respectively. These two figures mean that the amount of retransmitted data of Service A is more than that of ServiceB. The difference of the amount of retransmitted data causes the difference of transmitted data amount shown in Fig. 6. Let us consider the measured TCP segment loss rate that is derived from divisions of the number of retransmitted packets by that of all the transmitted packets; it is indicated in Fig. 9. From this figure, we can confirm that the measured packet loss rate is about the same as the one added by the network emulator.

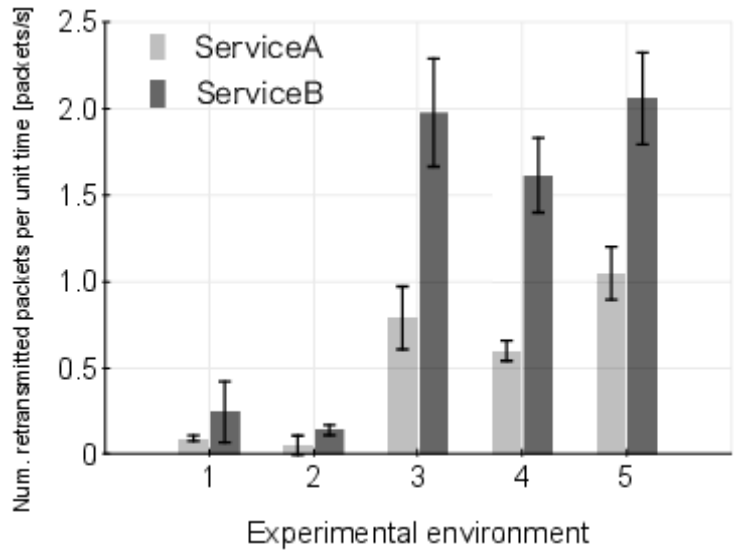

Figure 7. Number of the retransmitted packets per unit time.

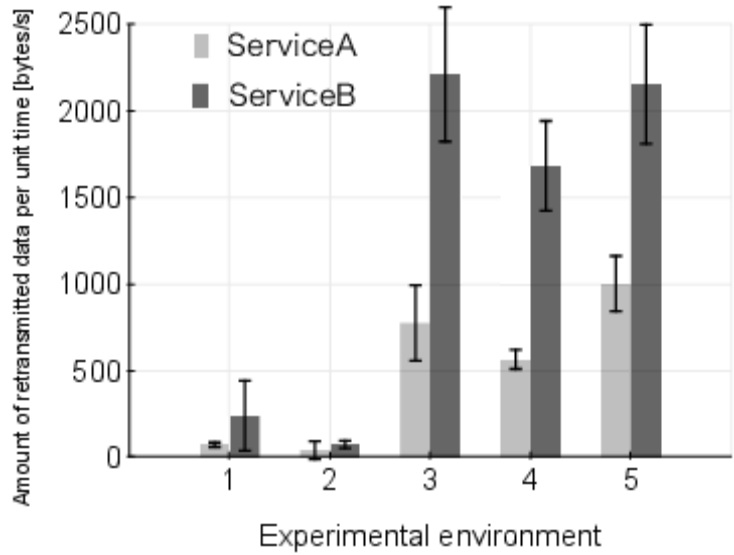

Figure 8. Amount of the retransmitted data per unit time.

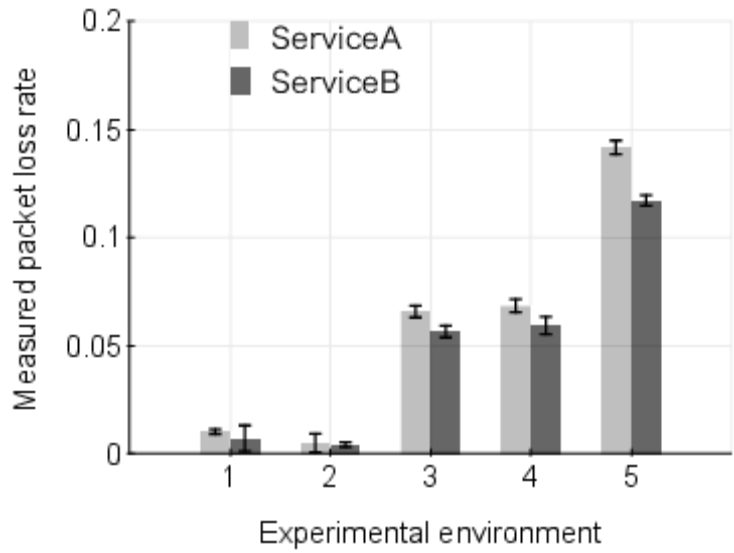

Figure 9. Measured packet loss rate.

### 6.2. Results of QoE assessment

For each experimental environment, Figs. 10, 11 and 12 plot the effectiveness, the efficiency and the satisfaction, respectively. These figures also show a 95% confidence interval for each plot.

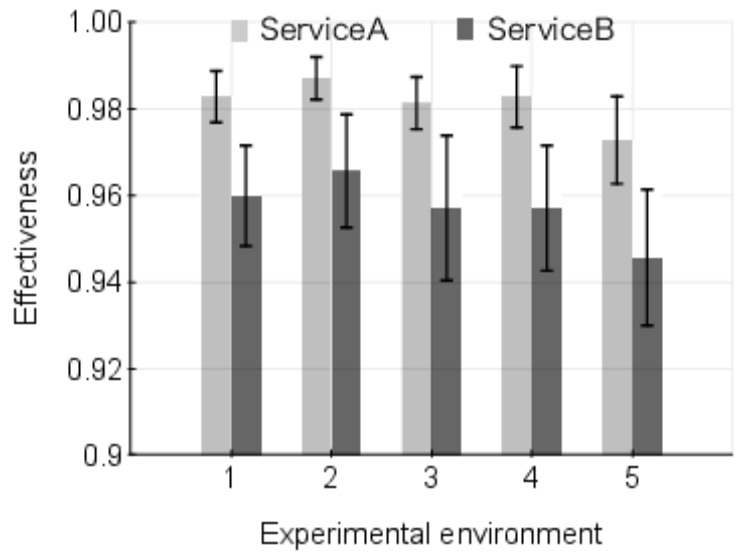

Figure 10. Measured effectiveness.

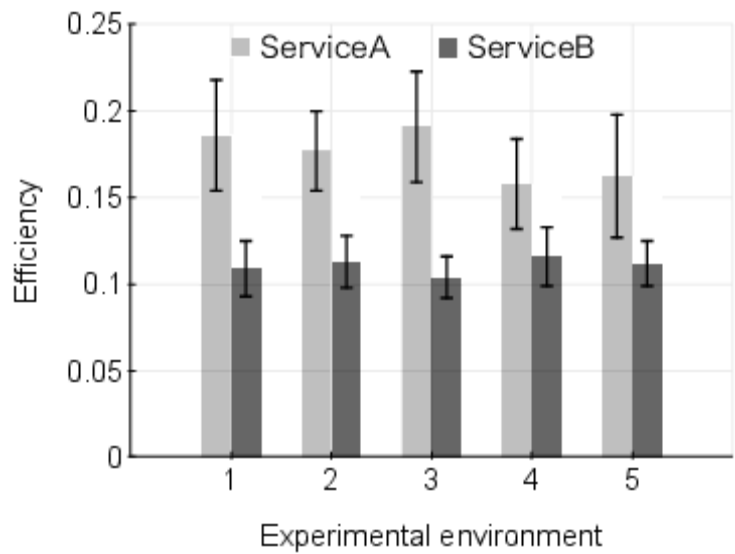

Figure 11. Measured efficiency.

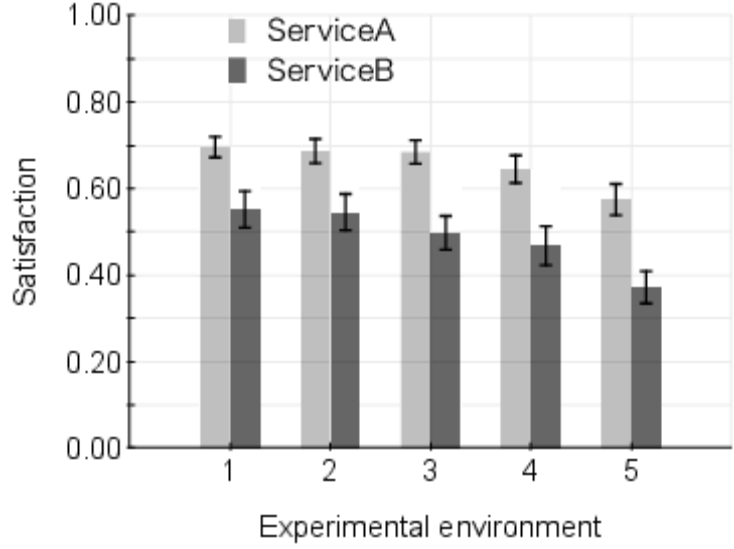

Figure 12. Measured satisfaction.

From Fig.10, we see that the effectiveness of ServiceA is better than that of ServiceB. We also find that, for both the services, the effectiveness decreases as the IP performance degrades. Fig. 11 indicates that the efficiency of ServiceA is slightly higher than that of ServiceB. However, we cannot confirm strong influence of the experimental environment on the efficiency. Fig. 12 shows that the satisfaction of ServiceA is slightly better than that of ServiceB. For both the services, the satisfaction also degrades because of IP performance degradation.

In order to clarify the relationship between QoE-Web and QoS quantitatively, we perform the multiple regression analysis; we treat the three measures of QoE-Web as the dependent variables and consider the metrics concerning QoS parameters as independent variables. Moreover, a dummy variable that denotes the service is added to the independent variables. For all the combinations of QoS parameters, we carry out the multiple regression analysis and choose a combination which makes the adjusted $R^2$ the highest. We show the results of the analysis in Eqs. (4) through (6). Note that, we remove the independent variables that are not statistically significant at 0.05.

$$\hat{E} = 0.985 - 0.00658T - 0.0196X \qquad (4)$$

$$\hat{H} = 0.1764 - 0.0623X \qquad (5)$$

$$\hat{S} = 0.693 - 0.00673T - 0.124X \qquad (6)$$

In these equations, $\hat{E}$、$\hat{H}$ and $\hat{S}$ are the estimate of the effectiveness, that of the efficiency and that of the satisfaction, respectively. $T$ and $X$ denote the retransmitted packets per unit time (packets/s) and the dummy variable, respectively. $X$ becomes 0 when the service is ServiceA and becomes 1 when the service is ServiceB. The adjusted $R^2$ of Eq. (4), that of Eq. (6) and that of Eq. (6) are 0.91, 0.93 and 0.95, respectively. According to Eqs. (4), (5) and (6), we plot the measured QoE-Web for the independent variable in Figs. 13, 14 and 15, respectively. Note that, in Eq. (5), the efficiency has no significant coefficient except for $X$. These figures also show the regression lines.

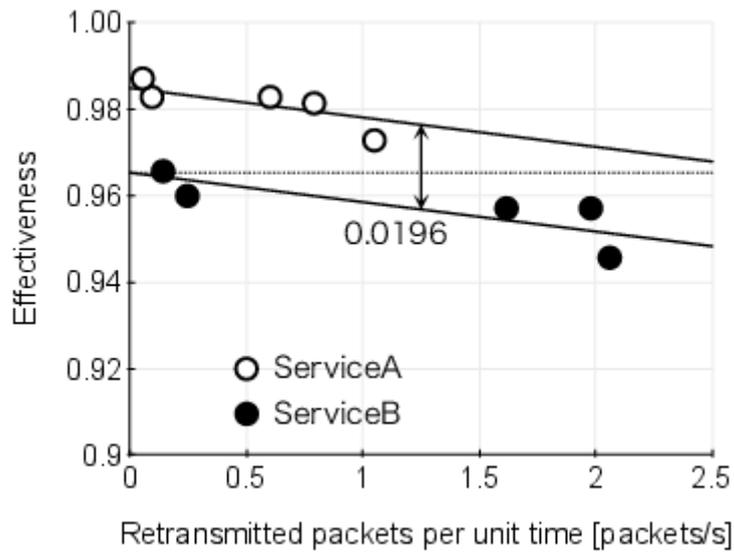

Figure 13. Effectiveness for retransmitted packets per unit.

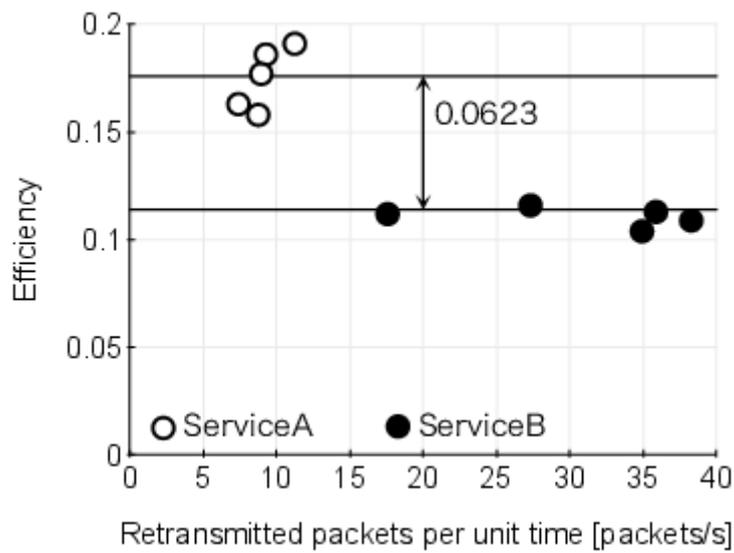

Figure 14. Efficiency for retransmitted packets per unit.

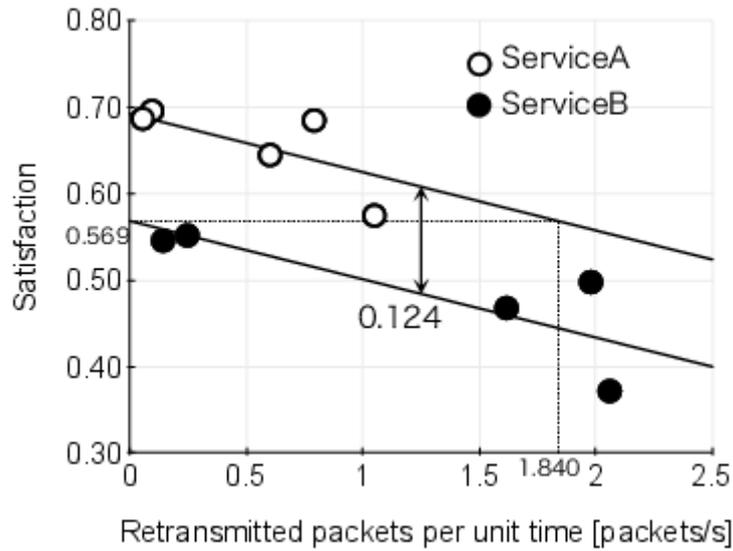

Figure 15. Satisfaction for retransmitted packets per unit.

From Eq. (4) and Fig. 13, we find that the effectiveness slightly degrades as the retransmitted packets increases. This means that the effectiveness relates with the Web design of the service more than the QoS. Since the coefficient of $X$ is 0.0196, the effectiveness for ServiceB is hardly above that for ServiceA. Equation (5) and Fig. 14 indicate that the efficiency affected only by the type of Web service.

Equation (6) and Fig. 15 shows that the satisfaction also deteriorates as the QoS degrades. The coefficient of $X$ is 0.124. Therefore, the satisfaction for ServiceA at 1.840 of $T$ becomes the same as that for ServiceB without any TCP retransmission. From these results, we find the followings. First, since the satisfaction deteriorates because of the QoS degradation, even if satisfaction of one service is higher than that of the other one in a good environment, the former can become lower than the latter according to difference between the environments of the two services. Second, indeed the QoS degradation degrades the effectiveness, but it hardly narrows the gap between the effectiveness of the two services. Third, the efficiency is affected only by the type of Web service, for example, Web designing, contents, and so on. As a result, we could quantitatively clarify the effect of QoS degradation on QoE-Web of the online shopping services. Consequently, we show that the proposed method of QoE-Web assessment is very useful to examine the relationship between QoE and QoS.

## 7. CONCLUSIONS

This paper proposed the method of clarifying the relationship between QoS degradation on QoE for a Web service and confirmed its effectiveness by experiment. In this experiment, we treat online Web services and consider the Web usability defined by ISO and the standard metrics defined by IETF as QoE-Web and QoS, respectively. Moreover, we examined the relationship between Web-QoE and QoS by the multiple regression analysis. The experiments utilized the two actual Web services. From the experimental results, we found the followings. The effectiveness is slightly affected by the QoS degradation. The efficiency and the satisfaction degrade extremely as the QoS deteriorates. We also clarified the difference between QoE of the two services quantitatively. From these results, it is significant to use Web usability for multidimensional Web-QoE evaluation and the proposed method is suitable for investigating the relationship between QoE and QoS. We have some issues for our future works. First, although this paper treats online shopping services, we would like to try other services. Second, we will tackle to use other measures as QoE-Web and QoS parameters.

**Authors**

**Daisuke Yamauchi** received the B. E. degree from Nagoya Institute of Technology in 2013. He is currently studying the master degree at Graduate School of Engineering, Nagoya Institute of Technology. His research interest is QoE assessment for Web services.

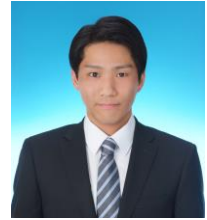

**Yoshihiro Ito** received the B.E., M.E., and Ph. D degrees from Nagoya Institute of Technology, Nagoya, Japan, in 1991, 1993, and 2002, respectively. From 1993 to 2001, he was with KDDI. IN 2001, he joined Nagoya Institute of Technology, in which he is now an Associate
Professor in the Department of Computer Science and Engineering, Graduate School of Engineering.
His research interests include multimedia communications over the Internet and QoE assessment. Dr. Ito is a member of IEEE and Information Processing Society of Japan.

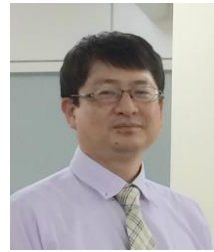